\documentclass[%
 reprint,
superscriptaddress,
 amsmath,amssymb,
 aps,
 prl,
]{revtex4-1}
\usepackage{dsfont}

\usepackage{nicefrac}
\usepackage{braket}
\usepackage{xcolor}
\usepackage{graphicx}
\usepackage{dcolumn}
\usepackage{bm}
\usepackage{hyperref}

\usepackage{algorithm}
\usepackage{algpseudocode}
\usepackage{algorithmicx}

\begin{document}

\preprint{APS/123-QED}

\title{Loss-tolerant quantum key distribution with a twist}

\author{J. Eli Bourassa}
 \email{bourassa@physics.utoronto.ca}
\affiliation{Department of Physics, University of Toronto, Toronto, Canada}

\author{Ignatius William Primaatmaja}
\affiliation{Centre for Quantum Technologies, National University of Singapore, Singapore}

\author{Charles Ci Wen Lim}
\email{charles.lim@nus.edu.sg}
\affiliation{Centre for Quantum Technologies, National University of Singapore, Singapore}
\affiliation{Department of Electrical and Computer Engineering, National University of Singapore, Singapore}

\author{Hoi-Kwong Lo}
\affiliation{Department of Physics, University of Toronto, Toronto, Canada}
\affiliation{Center for Quantum Information and Quantum Control, University of Toronto, Toronto, Canada}
\affiliation{Department of Electrical and Computer Engineering, University of Toronto, Toronto, Canada}
\affiliation{Department of Physics, University of Hong Kong, Hong Kong}

\date{\today}

\begin{abstract}
The security of measurement device-independent quantum key distribution (MDI QKD) relies on a thorough characterization of one's optical source output, especially any noise in the state preparation process. Here, we provide an extension of the loss-tolerant protocol [Phys. Rev. A 90, 052314 (2014)], a leading proof technique for analyzing the security of QKD, to MDI QKD protocols that employ mixed signal states. We first reframe the core of the proof technique, noting its generalization to treat $d$-dimensional signal encodings. Concentrating on the qubit signal state case, we find that the mixed states can be interpreted as providing Alice and Bob with a virtual shield system they can employ to reduce Eve's knowledge of the secret key. We then introduce a simple semidefinite programming method for optimizing the virtual operations they can perform on the shield system to yield a higher key rate, along with an example calculation of fundamentally achievable key rates in the case of random polarization modulation error. 
\end{abstract}

\maketitle

\section{Introduction}\label{sec:intro}

There has been a lot of interest in the subject of quantum hacking against practical QKD systems~\cite{ScaraniRMP2009,XuRMP2020}. In particular, single photon detectors have turned out to be the weakest link in the security of practical QKD. To completely short circuit all possible attacks on single photon detectors, the concept of measurement-device-independent QKD has been introduced and widely deployed. Measurement device-independent quantum key distribution (MDI QKD) protocols allow two distant parties, Alice and Bob, to distribute a shared, secret cryptographic key, even in the presence of an eavesdropper, Eve, who has complete control of their quantum channels and the measurement devices employed in the protocol \cite{Lo2012} (see also Ref.~\cite{Pirandola2012}). Typically, Alice and Bob prepare a set of signal states, send them to a central measurement node potentially controlled by Eve, which then makes an announcement based on a measurement it may or may not have faithfully executed. The cost of the information-theoretic security in such a setting is that Alice and Bob need to trust and characterize the optical sources they employ to send signals to the measurement devices. Thus, a proper understanding of the source features and flaws, and knowing how to account for them in a security proof is especially valuable for quantifying the key rates offered by an MDI protocol. 

In this work, we answer a seemingly simple question: how do you construct a security proof for an MDI QKD protocol that employs trusted, yet noisy -- i.e. mixed -- signal states? To clarify, protocols that employ the decoy state method \cite{Hwang2003,Lo2005} call for mixed optical states in the form of phase-randomized weak coherent pulses. However, in those protocols, the \textit{signal} states -- i.e. the single photon contributions -- from which the security of the key is derived are still often assumed to be pure. In this work, we specifically consider the issue of noisy signal states for MDI QKD protocols. We emphasize that having a consistent framework for optimally determining the security of QKD protocols employing mixed signal states from a trusted source is valuable since in any practical implementation of a protocol, even if the protocol calls for pure signal states, a realistic source will not be able to initialize a state with perfect purity; thus, we desire a security proof that can take into account the mixed nature of the signal states in an optimal manner, given that Eve may not hold the purification of the mixture. 

There are several leading proof techniques for handling state preparation errors in a QKD protocol. The first major analysis was performed by Gottesman, Lo, Lutkenhaus and Preskill \cite{GLLP2003}; however, they assumed pessimistically that Eve could amplify such noise to her benefit, and so the technique was not robust over long distances against e.g. coherent modulation errors. An improved technique was provided in the loss-tolerant protocol \cite{loss_tol}, which uses basis mismatch statistics to infer phase error rates that cannot be directly observed in the case state preparation is non-ideal. However, the technique leaves ambiguous how to treat \textit{mixed} signal states, a gap this work closes. A different extension of the loss tolerant protocol was considered in \cite{Pereira2019}; however, their focus was primarily on leaky sources, so their treatment of mixed states was analogous to \cite{loss_tol}. Another notable technique for characterizing security given pure qubit signal states is provided by \cite{PhysRevA.88.062322}; however, their technique for generalizing to mixed signal states uses a suboptimal approach of calculating the key rates for each of the pure states in the mixture, then averaging them, which will yield an equal or lower key rate than the key rate produced from the true average signal state. Lastly, an approach for finding a numerical lower bound on the Devetak-Winter secret key rate \cite{Devetak2005} for MDI-QKD protocols is provided by \cite{Coles2016,Winick2018reliablenumerical}; their technique is in principle extendable to noisy state preparation. In our work, we take a conceptually simpler strategy of directly optimizing the key rate formula from \cite{lo_sixstate}, which uses the bit, $x$ and $y$ phase errors of qubits in a virtual picture of the protocol.

In the case state preparation noise can be trusted and characterized, but perhaps not reduced, we provide here a simple analytical and numerical toolbox for calculating an optimal secret key rate. First, we provide a re-framing of the tilted four-state loss-tolerant protocol which provides a method for fixing Eve's degrees of freedom in the secret key rate \cite{loss_tol,LT_exp,LT_exp2}. However, as the signal states are mixed, the security also depends on how we treat the trusted noise in the signal state generation. Typically, the security of QKD is analyzed in terms of Alice and Bob's ability to virtually distill maximally entangled EPR pairs, since measurement of such pairs yields perfectly correlated keys, and by the monogamy of entanglement, the results cannot be correlated with anyone else, including Eve. However, it is known that a larger class of states known as private states \cite{twisted_eph,private_states,Renes_add_noise,private_states2} are fundamentally what is required to produce secret key. Formally, private states can be constructed from an EPR pair if Alice and Bob take ancillary shield systems they control, and apply a ``twisting" unitary operation between the EPR pair and the shields, the condition being that this twisting leave unaffected the measurement results that generate secret key. Since twisting does not change the key, private states can then be understood as deflecting some of Eve's attack on the systems that generate key to the shield systems. See Fig. \ref{fig:spaces} for a diagram of this concept.

In our technique, we show that the mixing noise of the signal states can be treated in a virtual picture as being equivalent to Alice and Bob employing shield systems that can be used to decrease Eve's knowledge of the key. Completely within this virtual picture, we can apply unitary twisting operations to the shields to decrease the phase errors of the protocol, increasing the secret key rate. We provide simple semi-definite programs to find the optimal twisting operations, yielding the optimal key rate under this framework. Semi-definite programming \cite{Boyd} has recently become a powerful tool for quantifying the asymptotic security of QKD protocols \cite{Coles2016,Winick2018reliablenumerical,Islam2018,Wang2019,Primaatmaja2019,Islam2019,Liu2019,Tan2019}. While private states have been of significant conceptual interest, as far as we are aware, this is the first application of private states in a practical QKD setting. Finally, we apply our technique to calculating fundamentally achievable key rates in an MDI QKD protocol with randomized modulation error in the state preparation procedure. We note that our technique is applicable to a general class of MDI QKD protocols in which Alice and Bob each employ four qubit signal states; as we will see, so long as their states do not fall in the same plane of the Bloch sphere (which is easy to impose in practice), they can be subject to general asymmetric preparation noise. Moreover, these signal states can be the single photon components of phase randomized coherent states in a decoy state protocol.

\section{Characterizing Eve's state}\label{sec:Eve}
We consider an MDI QKD protocol in which Alice and Bob each prepare four mixed qubit signal states $\{\rho_{A}^{i,x}\}$ and  $\{\sigma_{B}^{j,y}\}$, that they will send respectively with probabilities $p^{i,x}$ and $q^{j,y}$ to the central measurement node controlled by Eve. Here, $i=0,1$, $j=0,1$ index two sets of states Alice and Bob each independently choose from. When Alice and Bob choose $i,j=(0,0)$ these are the key generation states. All other settings $i,j$ correspond to test states used to constrain the phase errors. When $i,j=(0,0)$, $(x,y)$ correspond to their key bit values, while for other combinations of $(i,j)$, $(x,y)$ simply index which test signal states are being sent. Following the security proof of the loss tolerant protocol \cite{loss_tol}, we require that the sets of states $\{\rho_{A}^{i,x}\}$ and $\{\sigma_{B}^{j,y}\}$ each form a tetrahedron on the Bloch sphere, meaning the Bloch vectors cannot all lie in the same plane \cite{loss_tol}. In a decoy state protocol, these signal states correspond to the single photon components of phase-randomized weak coherent pulses; in the Supplementary Material, we provide steps for how to use our technique within a decoy state protocol in the asymptotic limit of an infinite number of decoys.

We now invoke that these are two-dimensional signal states, so each can be fully characterized with two orthonormal basis vectors, which we can take without loss of generality to be the polarization states $\ket{H},\ket{V}$:
\begin{equation}\label{eq:noisy}
    \rho_{A}^{i,x}\sigma_{B}^{j,y} = \sum_{\substack{m,m',\\n,n'=H}}^V c_{m,m'}^{i,x}d_{n,n'}^{j,y}\ket{m,n}\bra{m',n'}_{A,B} 
\end{equation}
Under unitary evolution, each of these basis vectors evolves to a (subnormalized) state in Eve's possession as well as a classical announcement, $z$, which we take to be pass or fail: $\ket{m,n}_{A,B}\rightarrow\sum_{z=P}^{F}\ket{e_{m,n}^z}_E\ket{z}_Z$. This process generalizes simply to multiple announcement events, such as which Bell state Eve claims to have detected.

The probabilities that Eve announces a round successfully passed as a function of the signal states sent $p_{det}^{i,j,x,y} = p(z=P|i,j,x,y)$, provide constraints on the inner product of Eve's vectors $\braket{e^P_{m',n'}|e^P_{m,n}}_E$:
\begin{equation}\label{eq:pdet}
    p_{det}^{i,j,x,y} = p^{i,x} q^{j,y}\sum_{\substack{m,m',\\n,n'=H}}^{V} c_{m,m'}^{i,x} d_{n,n'}^{j,y}\braket{e^P_{m',n'}|e^P_{m,n}}_E
\end{equation}
$p_{det}^{i,j,x,y}$ are observable quantities in the protocol, and they can also be used to directly calculate some quantities required for the secret key rate formula, such as the detection probability in the key basis, $p_{det}^{0,0} = \sum_{x,y}p_{det}^{0,0,x,y}$, and the bit error rate $e_Z = (p_{det}^{0,0,0,1} + p_{det}^{0,0,1,0})/p_{det}^{0,0}$, where we have taken $\ket{\Phi^{+}}$ to be the target Bell state that Alice and Bob wish to distill in a virtual picture we describe in the next Section. 

For qubit signal states there are sixteen combinations of $(m,n,m',n')$, which we flatten to one label $s=1,...,16$. Thus, we can define $\vec{e}_s = \braket{e^P_{m',n'}|e^P_{m,n}}_E$, the vectorized form of the Gramian matrix of Eve's states associated with a passing announcement. Solving for $\vec{e}$ means we can then calculate any objective function of $\braket{e^P_{m',n'}|e^P_{m,n}}_E$, including all the phase error rates in the six-state protocol key rate formula \cite{loss_tol,lo_sixstate,Renner_thesis,RevModPhys.81.1301}, even though we are only using four states, since the four states are chosen to provide complete characterization of Eve's strategy. Additionally, we see that with Alice and Bob sending four states each, $(i,j,x,y)$ also takes on sixteen combinations, which we can label with $t=1,...,16$. Hence, we can define a vector containing all the successful detection probabilities $(\vec{p}_{det})_t = p_{det}^{i,j,x,y}$, and a matrix dependent on the initial states from Eq. \ref{eq:noisy} used in the protocol $\hat{\gamma}_{ts} = p^{i,x} q^{j,y} c_{m,m'}^{i,x} d_{n,n'}^{j,y}$.

We see that Eq. \ref{eq:pdet} can be written compactly as:
\begin{equation}\label{eq:solving_Eve}
    \vec{p}_{det} = \hat{\gamma} \vec{e} \implies \vec{e} = \hat{\gamma}^{-1}\vec{p}_{det}
\end{equation}
where we additionally see that as long as $\hat{\gamma}^{-1}$ exists, then we can exactly solve for $\vec{e}$. We note the basis ordering of $\hat{\gamma}$ can be chosen so that its rows are the tensor product of the vectorized forms of the probability-weighted signal states $\text{vec}(p^{i,x} \rho_A^{i,x})^T\otimes\text{vec}(q^{j,y}\sigma_B^{j,y})^T$. Since invertibility of $\hat{\gamma}$ is equivalent to its rows being linearly independent, this provides a clear minimum requirement for state preparation, namely that the vectorized forms of the states which Alice and Bob use in the protocol must be linearly independent. In the Supplementary Material, we show that meeting this condition is equivalent to sending four states that form a tetrahedron on the Bloch sphere, as found in the original security proof of the tilted four-state loss tolerant protocol \cite{loss_tol}.

\textbf{Remark: }Having reframed the security proof from the loss-tolerant protocol in this form, we observe that the process from Eq. \ref{eq:solving_Eve} can also generalize straightforwardly to MDI QKD protocols employing discrete-variable high-dimensional degrees of freedom, such as those employing orbital angular momentum \cite{Sit2017,Wang2015,Wang2019OAM} or timebin encodings \cite{Tang2014,Islam2019}. For Alice and Bob each sending $d$-dimensional systems, the Gramian matrix $\braket{e^P_{m',n'}|e^P_{m,n}}_E$ of Eve's states contains in general $d^4$ elements. Thus, Alice and Bob can each prepare $d^2$ states within the $d$-dimensional space, that will in turn yield $d^4$ observable detection probabilities. They can prepare their states subject to $\text{vec}(p^{i,x} \rho_A^{i,x})\otimes\text{vec}(q^{j,y}\sigma_B^{j,y})$ being linearly independent, where $(i,j,x,y)$ has $d^4$ possibilities. This condition of linear independence for the vectorized density matrices is a much less stringent condition to satisfy for high-dimensional protocols than e.g. employing eigenstates of a sufficient number of mutually unbiased bases \cite{Mafu2013,Cozzolino2019}, while still allowing for complete characterization of the parameters in the high-dimensional secret key rate formula that are dependent on Eve's system \cite{Sheridan2010}.

\section{Optimal choice of virtual protocol}\label{sec:virtual}

\begin{figure}
    \begin{center}
        \includegraphics[width = 6 cm] {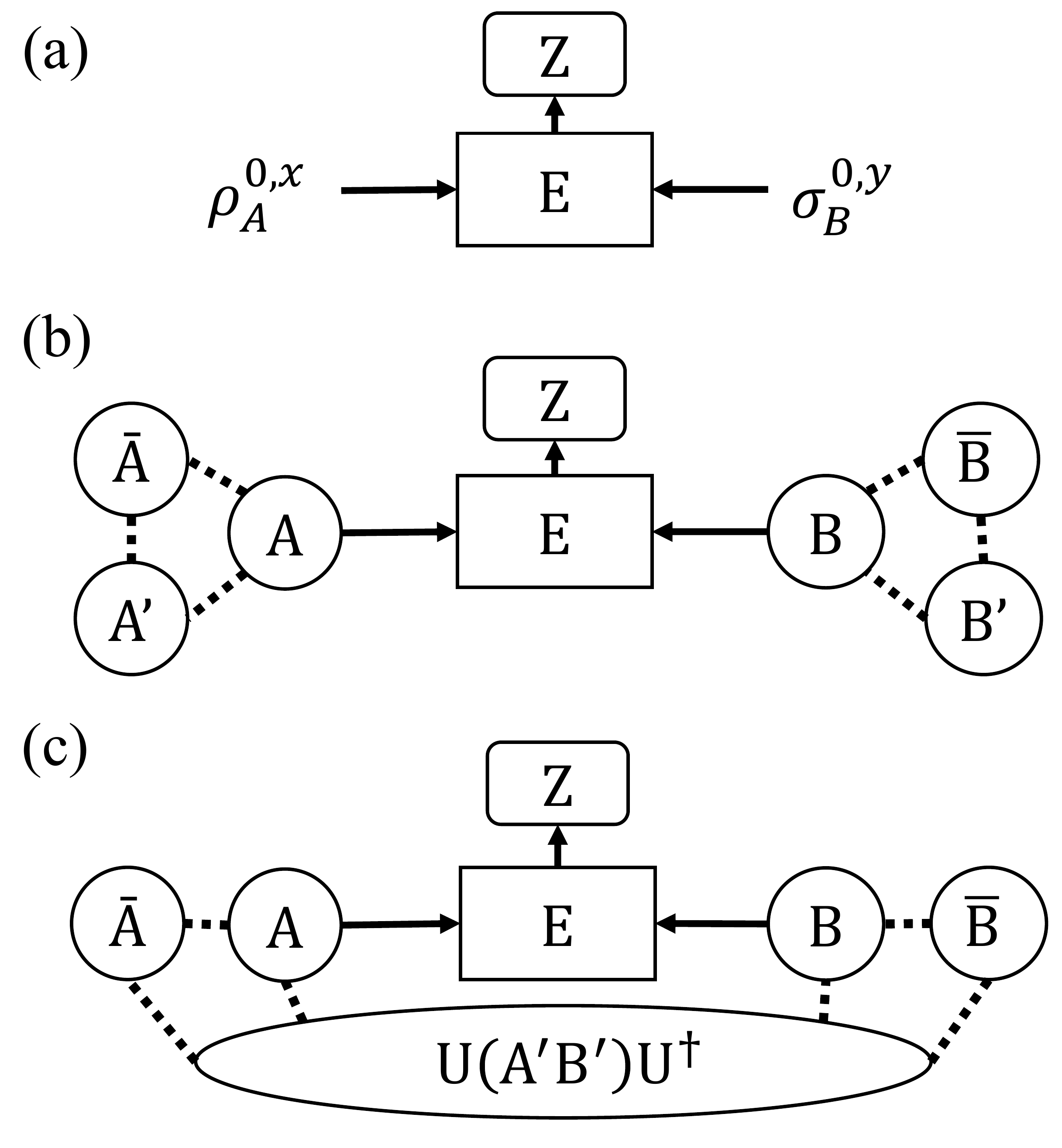}
    \end{center}
    \caption{(a) A real MDI QKD protocol: Alice and Bob each prepare mixed states associated with bit $(x,y)$ and basis $(i,j)$ values. They send their states to a central node controlled by Eve, who makes an announcement Z. (b) A virtual version of the key generation states in the protocol: in a purified picture, Alice and Bob's mixed signal states are entangled with virtual qubits $\bar{A}\bar{B}$ which coherently store the bit values $(x,y)$. Measurement of $\bar{A}\bar{B}$ in the computational basis yields the raw keys. The $AB$ systems are additionally purified by the $A'B'$ systems to account for trusted noise in the source. Only the $A,B$ systems are sent to Eve. (c) An alternative virtual purification: all purifications are related by unitary operations applied to, in general, a joint purifying ancilla, yielding private states in $\bar{A}\bar{B}A'B'$. These ``twisting" operations can optimally boost the secret key rate as they can modify the phase error rates which Alice and Bob need to estimate. In (a)--(c), the signal states sent and the observed protocol statistics (detection and bit error rates) are the same.}
    \label{fig:spaces}
\end{figure}

Having characterized Eve's Gramian matrix entirely from observable parameters in the protocol, we now move to a virtual picture for the key generation signal states to calculate the remaining parameters of the secret key rate. In this virtual picture, which is depicted in Fig. \ref{fig:spaces}, the states of systems $A,B$ from Eq. \ref{eq:noisy} are entangled with virtual qubits $\bar{A},\bar{B}$ that Alice and Bob keep in their lab \cite{loss_tol}. Importantly, since these signal states are mixed, we require additional purifying ancillary systems $A'B'$. We assume that the sources of noise are confined to Alice and Bob's labs, meaning we can trust Eve does not have access to manipulate $A'B'$. The mixedness of the signal states then results in an effective virtual shield Alice and Bob can use to minimize Eve's knowledge of the secret key.

The key generation states, $p^{0,x} q^{0,y}\rho_{A}^{0,x}\sigma_{B}^{0,y}$ can be considered virtually as a state of the form:
\begin{equation}\label{eq:pure}
    \ket{\zeta} = \sum_{x,y}|x,y\rangle_{\bar{A}\bar{B}}\sum_{m,n = H}^{V} |\gamma^{x,y}_{m,n}\rangle_{A'B'}|m,n\rangle_{AB}
\end{equation}
where we have constraints from the states in Eq. \ref{eq:noisy}:
\begin{equation}\label{eq:constraint}
    \braket{\gamma^{x,y}_{m',n'}|\gamma^{x,y}_{m,n}}_{A'B'} =  p^{0,x} q^{0,y} c_{m,m'}^{0,x} d_{n,n'}^{0,y} = \hat{\gamma}_{ts}
\end{equation}
since to generate key, Alice and Bob measure $\bar{A}\bar{B}$ in the computational basis. The crucial point is that this purification is not unique \cite{NielsenChuang}, and so we have freedom to choose the virtual picture that yields the optimal key rate. Since Eve does not have access to $A'B'$, any purification will yield a suitable lower bound on the key rate, but we will show how to choose the optimal purification with simple semidefinite programs.

We can parametrize all purifications using twisting unitary operations \cite{twisted_eph,private_states,Renes_add_noise,private_states2} applied to the virtual ancillary systems in $\ket{\zeta}$:
\begin{equation}\label{eq:twist_op}
    U_{\bar{A}\bar{B}A'B'} = \sum_{x,y=0}^{1}|x,y\rangle\langle x,y|_{\bar{A}\bar{B}}\otimes U_{A'B'}^{x,y}
\end{equation}
Such an operation is entirely virtual, so it can be nonlocal in general and never needs to be executed in the real protocol. Twisting does not affect any of the real observed detection probabilities, which correspond to Alice and Bob first projecting $\bar{A}\bar{B}$ in the computational basis, as we show in the Supplementary Material. Moreover, since only the $A,B$ portion of $\ket{\zeta}$ evolves unitarily to $E,Z$, the twisting operation need not be fixed from the beginning of the protocol, and its choice can and should be informed by the statistics of the protocol. We next show exactly how these twisting operations affect the secret key formula.

To quantify the security of the protocol, we employ the key rate formula from the six-state protocol \cite{lo_sixstate,RevModPhys.81.1301,Renner_thesis}, noting, however, that our protocol employs only four states:
\begin{equation}\label{eq:six_state}
\begin{split}
        R = p_{det}^{0,0}\Bigg(&1-h_2(e_{Z})- e_Z h_2\left[\frac{1+(e_X-e_Y)/e_Z}{2}\right] \\  &  - (1-e_Z)h_2\left[\frac{1-(e_X+e_Y+e_Z)/2}{1-e_Z}\right]\Bigg)
\end{split}
\end{equation}
where $h_2(\cdot)$ is the binary entropy function, and $e_{X}$ and $e_{Y}$ are the phase error rates of the virtual qubits $\bar{A}\bar{B}$ in the $X$ and $Y$ Pauli bases. These can be understood as the probability of the virtual qubits being projected into the incorrect Bell states given a passing announcement from Eve:
\begin{equation}\label{eq:exey}
    \begin{split}
    e_{X}  &= \frac{\langle\Gamma| \left[\right.(|\Psi^-\rangle\langle \Psi^-|
    +|\Phi^-\rangle\langle \Phi^-|)_{\bar{A}\bar{B}}\otimes|P\rangle\langle P|_{Z} \left.\right] |\Gamma\rangle}{\langle\Gamma| (|P\rangle\langle P|_{Z}) |\Gamma\rangle}\\
    e_{Y}  &= \frac{\langle\Gamma| \left[\right.(|\Psi^+\rangle\langle \Psi^+|
    +|\Phi^-\rangle\langle \Phi^-|)_{\bar{A}\bar{B}}\otimes|P\rangle\langle P|_{Z} \left.\right] |\Gamma\rangle}{\langle\Gamma| (|P\rangle\langle P|_{Z}) |\Gamma\rangle}\\
\end{split}
\end{equation}
Here, $\ket{\Gamma}$ denotes the joint state between $\bar{A}\bar{B}A'B'EZ$ after the $AB$ portion of twisted purified state $U_{\bar{A}\bar{B}A'B'}\ket{\zeta}$ is sent to Eve. Note as well $\langle\Gamma| (|P\rangle\langle P|_{Z}) |\Gamma\rangle = p_{det}^{0,0}$. The six-state protocol key rate provides generally higher key rates than the Shor-Preskill key rate \cite{Shor2001} because it takes into account correlations between the bit and phase error patterns. 

Employing Eq. \ref{eq:pure}, the form of the twisting operation in Eq. \ref{eq:twist_op}, and the unitary evolution $\ket{m,n}_{A,B}\rightarrow\sum_{z=P}^{F}\ket{e_{m,n}^z}_E\ket{z}_Z$, we find that $e_X$ and $e_Y$ are linear functions with respect to the elements of Eve's Gramian matrix $\braket{e^P_{m',n'}|e^P_{m,n}}_E$, which are already known from Eq. \ref{eq:solving_Eve}. Additionally, the phase errors are linear with respect to matrix elements $\braket{\gamma^{x',y'}_{m',n'}|U_{A'B'}^{x',y'\,\dagger}U_{A'B'}^{x,y}|\gamma^{x,y}_{m,n}}_{A'B'}$, which are functions of the twisting operation we control. We see these elements form the Gramian matrix of the twisted ancillary system states. Since our task is to modify the twisting operation to boost the key rate, these elements form the optimization variables of our problem. Their only constraint is provided by Eq. \ref{eq:constraint}, since by construction when $x=x',y=y'$, the twisting operations cancel to not affect the form of the real protocol signal states.

Of course, the binary entropy function in the key rate formula from Eq. \ref{eq:six_state} breaks the linearity of the key rate optimization with respect to $e_X$ and $e_Y$; however, the linear combinations $e_{\pm } = e_X \pm e_Y$ can instead be chosen as the objective functions of the optimization, since they are still linear with respect to $\braket{\gamma^{x',y'}_{m',n'}|U_{A'B'}^{x',y'\,\dagger}U_{A'B'}^{x,y}|\gamma^{x,y}_{m,n}}_{A'B'}$. Moreover, we find that $e_+$ only depends on $U_+ = U^{0,0\,\dagger}_{A'B'}U^{1,1}_{A'B'}$ and $e_-(U)$ only depends on $U_- = U^{0,1\,\dagger}_{A'B'}U^{1,0}_{A'B'}$. Intuitively, we have such a dependence since $e_-$ involves the Bell states that underwent a bit flip, so Alice and Bob's bit values will be (0,1) and (1,0), and only those twisting unitaries will be used. Similarly $e_+$ reflects Bell states that did not undergo a bit flip, so twisting will only involve (0,0) and (1,1). Since the unitaries $\{U^{0,1}_{A'B'},U^{1,0}_{A'B'},U^{0,0}_{A'B'},U^{1,1}_{A'B'}\}$ can be defined independently of each other, the optimization of $e_+$ can be decoupled from the optimization of $e_-$, so we can overcome the nonlinearity introduced by the binary entropy function in Eq. \ref{eq:six_state} by taking advantage of the monotonicity of $h_2(\cdot)$ and optimizing the arguments $e_{\pm}$ of the function instead.

Taking stock, we have two independent objective functions $e_{\pm}$, which are linear with respect to our optimization variables $\braket{\gamma^{x',y'}_{m',n'}|U_{A'B'}^{x',y'\,\dagger}U_{A'B'}^{x,y}|\gamma^{x,y}_{m,n}}_{A'B'}$. These variables form a Gramian matrix, which is a positive semidefinite matrix by construction, and they are additionally subject to linear constraints from Eq. \ref{eq:constraint}. Thus, these optimization problems take the form of semidefinite programs which can be solved numerically on a standard laptop in a few seconds using available packages for Python \cite{cvxpy,cvxpy_rewriting} or Matlab \cite{cvx}. While previous literature on twisting operations had noted the opportunity for optimizing $U$ \cite{private_states2}, no explicit procedure was constructed. Here, we have closed this gap, increasing the practicality of utilizing a virtual twisting operation as a step in the security proof. In the Supplementary Material, we provide complete details for framing the problem in terms of semidefinite programs, along with the expressions for the objective functions $e_{\pm}$ and their dependence on the Gramian matrix of the $A'B'$ system.

\section{Key rate results}\label{sec:results}

We emphasize again that the only requirement for applying our technique is that Alice and Bob's initial qubit signal states cannot fall in the same plane of the Bloch sphere, which is easy to satisfy in practice. Otherwise, our technique can handle quite general noisy state preparation: Alice and Bob need not prepare the same sets of states; they can send states with different probabilities; and, the noise channel applied to each state can be dependent on the state. 

As a study of fundamentally achievable key rates, we consider the following two-parameter $(\delta,p)$-model for the initial states. We suppose Alice and Bob attempt to prepare the states $\{\ket{H},\ket{V},\nicefrac{\ket{H}+\ket{V}}{\sqrt{2}},\nicefrac{\ket{H}-i\ket{V}}{\sqrt{2}}\}$; however, each state is subject to a modulation error which we treat as a random variable. The error and its distribution can differ depending on the state. Given a distribution for the modulation error on the surface of the Bloch sphere, the resulting average states can be treated as having a constant coherent modulation error, i.e. a constant offset angle from the ideal state, parametrized by $\delta$, as well as a depolarization noise parametrized by $p$, which shortens the Bloch vector and introduces incoherent mixing to the states. For exact definitions of the signal states, see the Supplementary Material. For the case $p=0$, we do not expect any improvement in our key rate over the standard loss tolerant protocol, since there would be no mixing and thus no virtual ancillary system on which to apply a twisting operation. 

\begin{figure}
    \begin{center}
        \includegraphics[width = \linewidth] {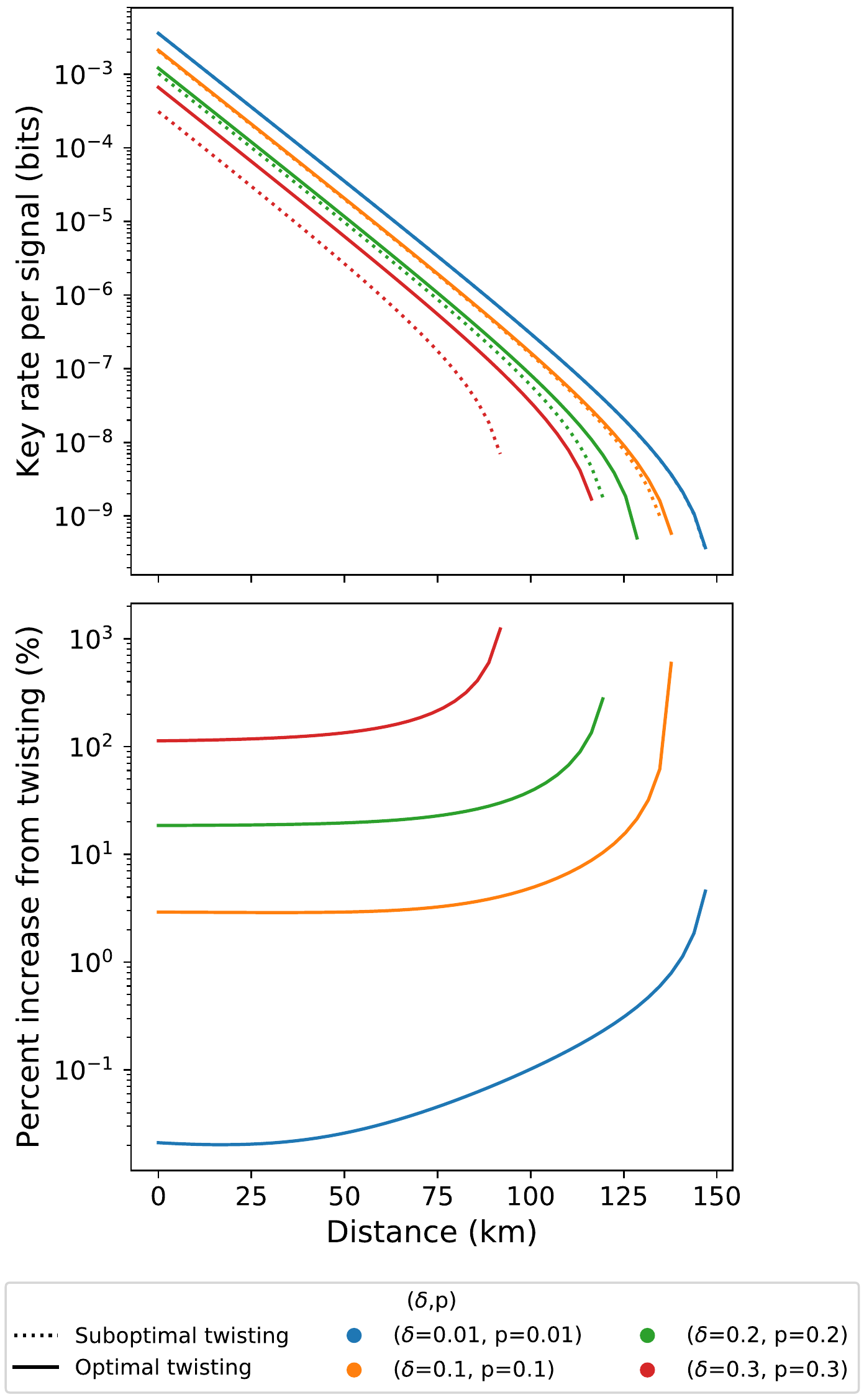}
    \end{center}
    \caption{Top: Key rate vs. Alice-Charlie distance for various values of modulation error and depolarizing noise $(\delta,p)$ (same color indicates same model parameters). The dotted lines are the results from a naive, suboptimal purification, while the solid line indicates our optimized key rate over all virtual twisting operations. Bottom: For each pair $(\delta,p)$, the percentage increase offered by optimizing over twisting operations. We see our technique is best suited to states with more preparation noise sent over longer distances. We assume a single photon source and symmetric distances from Alice and Bob to Charlie.}
    \label{fig:results}
\end{figure}

In Fig. \ref{fig:results}, we plot the asymptotic key rate found using our technique as a function of distance for various pairs $(\delta,p)$. We assume a Bell state detection scheme similar to \cite{Lo2012}, with overall detection efficiency of 50\%, a dark count probability of $10^{-5}$ per pulse per detector, loss in fiber of 0.2 dB/km, and error correction efficiency of 1. For comparison with the key rate produced with our optimization, we plot the key rate calculated using a suboptimal purification, which was constructed by simply diagonalizing Alice and Bob's signal states and having the ancillary systems index the eigenvalues in decreasing order. 

We find that our technique provides a modest increase over the ``naive" purification one could have chosen, our technique's advantages being most significant as the depolarizing noise gets stronger (making the initial states more mixed), and at longer distances when the untrusted channel noises (loss and dark counts) accrue. Additionally, we see a better key rate can be produced by reducing state preparation noise; however, once one has improved as best possible, our technique provides an optimized key rate given that level of noise. That is, our technique provides confidence that one has optimized over all possible ancillary states of the purification that are consistent with the protocol statistics without worry that one has chosen a pessimistic virtual picture.

\section{Conclusion}
We have presented an extension of the proof technique from \cite{loss_tol} to quantify the security of MDI QKD protocols that employ general noisy qubit signal states. We first reframed the analytical technique used for characterizing the parameters in the secret key rate that depend on Eve's system, noting that this new approach lends itself clearly to a generalization for higher-dimensional signal states. Next, we observed that employing trusted but mixed signal states means Alice and Bob have not a single but a set of virtual pictures they can use to analyze security in their protocol; we observed this was equivalent to Alice and Bob employing a \textit{virtual} shield system onto which they can apply \textit{virtual} twisting operations to minimize Eve's knowledge of the key \cite{private_states,twisted_eph}. Finally, we provided a simple numerical technique leveraging semidefinite programming to optimize over all twisting operations to optimize the six-state protocol secret key rate formula, examining the implications for state preparation subject to random modulation error.

\section{Acknowledgements}
We thank Ilan Tzitrin, Wenyuan Wang, Thomas van Himbeeck, Emilien Lavie, and Koon Tong Goh for useful discussions. We thank NSERC, CFI, ORF, MITACS, Royal Bank of Canada, Huawei Technology Canada, and HKU start-up grants for funding support. J.E.B. is supported by an Ontario Graduate Scholarship and the Lachlan Gilchrist Fellowship. I.W.P. is supported by National Research Foundation and the Ministry of Education, Singapore, under the Research Centres of Excellence programme. C.L. acknowledges support by the National Research Foundation (NRF) Singapore, under its NRF Fellowship programme (NRFF11-2019-0001) and Quantum Engineering Programme (QEP-P2).

\bibliography{biblio}

\end{document}


\title{Supplementary materials: Loss-tolerant quantum key distribution with a twist}

\author{J. Eli Bourassa}
 \email{bourassa@physics.utoronto.ca}
\affiliation{Department of Physics, University of Toronto, Toronto, Canada}
\author{Ignatius William Primaatmaja}
\affiliation{Centre for Quantum Technologies, National University of Singapore, Singapore}

\author{Charles Ci Wen Lim}
\email{charles.lim@nus.edu.sg}
\affiliation{Centre for Quantum Technologies, National University of Singapore, Singapore}
\affiliation{Department of Electrical and Computer Engineering, National University of Singapore, Singapore}

\author{Hoi-Kwong Lo}
\affiliation{Department of Physics, University of Toronto, Toronto, Canada}
\affiliation{Center for Quantum Information and Quantum Control, University of Toronto, Toronto, Canada}
\affiliation{Department of Electrical and Computer Engineering, University of Toronto, Toronto, Canada}
\affiliation{Department of Physics, University of Hong Kong, Hong Kong}
\date{\today}
\maketitle

\section{Embedding our technique within a decoy state protocol}

In the main text, we showed that, using two independent semidefinite programs (SDP), we can optimize the secret key rate of a loss-tolerant protocol where the signal states are two-dimensional mixed states. In the context of quantum key distribution, those states are normally encoded in the mode (such as polarization or time-bin) of single photons. However, in practice, many protocols employ weak coherent pulses with decoy states \cite{Hwang2003,Lo2005}. In this section, we outline how we can embed our technique in a loss-tolerant protocol which uses decoy states. In this work, we will consider a protocol with infinite number of decoy states as in \cite{loss_tol}. We leave the case where a finite number of decoy states are used for future work. Like in the main text, we work in the asymptotic limit where we can ignore finite key effects.

Recall from the main text that to use our technique, we need two pieces of information: before sending optical signals to Eve, we need the density matrices of the qubit signal states (the single photon component), and then once the optical signals are sent, we need the probability of successful Bell state measurement given the states that Alice and Bob chose. As such, to apply our technique when phase-randomised weak coherent pulses are used together with a decoy state protocol, we need to calculate the state of the single photon component of the optical signal as well as the probability of successful Bell state measurement given that Alice and Bob send the corresponding single photon signals. In the literature, that conditional probability is often referred to as the single photon yield, denoted by $Y_{11}^{i,j,x,y}$.

To obtain the single photon component of the signals, we can simply project the coherent signal states to their single photon components. Suppose that Alice and Bob prepare the phase-randomised coherent state $\tilde{\rho}_{A}^{i,x,\mu_A}$ and $\tilde{\sigma}_{B}^{j,y,\mu_B}$ with intensity $\mu_A$ and $\mu_B$ respectively, the single photon components of those states can be easily obtained by performing the following projection and then normalizing the resulting states
\begin{equation}
    \begin{split}
        \rho_A^{i,x} &= \frac{\Big(\ketbra{0}{0}_{H_A} \otimes \ketbra{1}{1}_{V_A} + \ketbra{1}{1}_{H_A} \otimes \ketbra{0}{0}_{V_A} \Big) \tilde{\rho}_{A}^{i,x,\mu_A} \Big(\ketbra{0}{0}_{H_A} \otimes \ketbra{1}{1}_{V_A} + \ketbra{1}{1}_{H_A} \otimes \ketbra{0}{0}_{V_A} \Big)}{e^{-\mu_A}\mu_A} \\
        \sigma_B^{j,y} &= \frac{\Big(\ketbra{0}{0}_{H_B} \otimes \ketbra{1}{1}_{V_B} + \ketbra{1}{1}_{H_B} \otimes \ketbra{0}{0}_{V_B} \Big) \tilde{\sigma}_{B}^{j,y,\mu_B} \Big(\ketbra{0}{0}_{H_B} \otimes \ketbra{1}{1}_{V_B} + \ketbra{1}{1}_{H_B} \otimes \ketbra{0}{0}_{V_B} \Big)}{e^{-\mu_B}\mu_B}
    \end{split}
\end{equation}
where $\ket{0}_m$ and $\ket{1}_m$ are the vacuum and single photon states in mode $m$ respectively. 

Hence, it is important that we characterize the sources of each legitimate party before performing the protocol. Ideally, this should be done by performing tomography on the signal states $\tilde{\rho}_{A}^{i,x,\mu_A}$ and $\tilde{\sigma}_{B}^{j,y,\mu_B}$. Alternatively, one can have a model for the source, taking into account the finite precision and randomness in the modulation of the signal states. Once we have the single photon component of the signal states (i.e. $\rho_A^{i,x}$ and $\sigma_B^{j,y}$ in Eq. (1) of the main text), we can use them to construct the $\hat{\gamma}$ matrix in Eq. (3) of the main text, and to impose the constraints on the ancillary systems $A'B'$ as described in Eq. (5) of the main text .

On the other hand, from the parameter estimation step of the protocol, we can estimate the gain $Q_{\mu_A,\mu_B}^{i,j,x,y}$ for each choice of states $(i,x)$ and $(j,y)$ and intensities $\mu_A, \mu_B$. When using infinite number of decoy states, Alice and Bob can determine the values of the single photon yield $Y_{11}^{i,j,x,y}$ exactly for all $i,j,x,y$. Once the values of $Y_{11}^{i,j,x,y}$ are obtained, we can replace the $p_{det}^{i,j,x,y}$ with $Y_{11}^{i,j,x,y}$ in Eq. (3) of the main text and then proceed with our method.

\section{Relationship between the invertibility of $\hat{\gamma}$ and the states in the Bloch sphere forming a tetrahedron}
In the main text, we demonstrated that we could solve for the elements of Eve's Gramian matrix using the equation:
\begin{equation}\label{eq:solving_Eve}
    \vec{p}_{det} = \hat{\gamma} \vec{e} \implies \vec{e} = \hat{\gamma}^{-1}\vec{p}_{det}
\end{equation}
where $\vec{e}_s = \braket{e_{m',n'}^P|e_{m,n}^P}_E$ are the elements of the vectorized form of the Gramian matrix of Eve's states associated with a passing announcement. $(\vec{p}_{det})_t = p_{det}^{i,j,x,y}$ form a vector containing all the successful detection probabilities, and $\hat{\gamma}_{ts} = p^{i,x} q^{j,y} c_{m,m'}^{i,x} d_{n,n'}^{j,y}$ form a matrix dependent on the initial states used in the protocol which were taken to be:
\begin{equation}\label{eq:noisy}
    \rho_{A}^{i,x}\sigma_{B}^{j,y} = \sum_{\substack{m,m',\\n,n'=H}}^V c_{m,m'}^{i,x}d_{n,n'}^{j,y}\ket{m,n}\bra{m',n'}_{A,B} 
\end{equation}

\begin{figure}
    \begin{center}
        \includegraphics[width = 12cm] {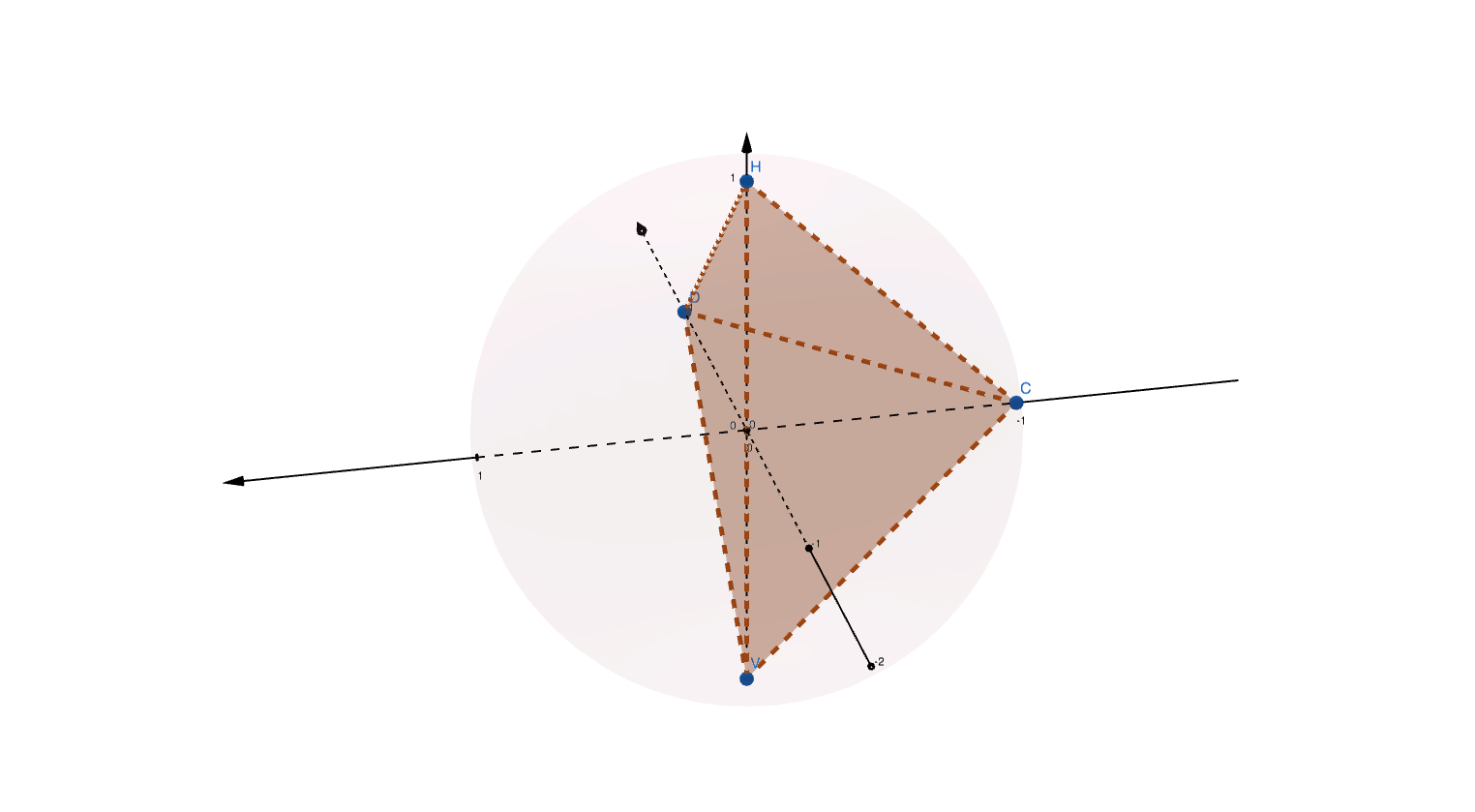}
    \end{center}
    \caption{A tetrahedron in the Bloch sphere representing qubits encoded in horizontal, vertical, diagonal and clockwise circularly polarized single photons. A tetrahedron is formed so long as the states don't all fall in the same plane.}
    \label{fig: tetra}
\end{figure}

Here we show that the invertibility of $\hat{\gamma}$ is equivalent to the condition in the loss tolerant protocol \cite{loss_tol} that Alice and Bob each need to choose four signal states that form a tetrahedron in the Bloch sphere, as shown in Fig. \ref{fig: tetra}. 

We begin by noting that:
\begin{equation}\label{eq:vec_AB}
 \hat{\gamma}_{ts} = p^{i,x} q^{j,y} c_{m,m'}^{i,x} d_{n,n'}^{j,y} = p^{i,x} q^{j,y}\braket{m,n|\rho_A^{i,x}\sigma_B^{j,y}|m',n'}_{A,B}
\end{equation}
meaning we can always choose the basis ordering of $\hat{\gamma}$ so that its rows are $\text{vec}(p^{i,x} \rho_A^{i,x})^T\otimes\text{vec}(q^{j,y}\sigma_B^{j,y})^T$, the tensor product of the vectorized forms of the probability-weighted signal states. Invertibility of $\hat{\gamma}$ is equivalent to showing its rows are linearly independent. Since all the row vectors have tensor product form, we just need to show that the $\{\text{vec}(p^{i,x} \rho_A^{i,x})\}$ and the $\{\text{vec}(q^{j,y}\sigma_B^{j,y})\}$ each form sets of linearly independent vectors.

Next, we recall that four states forming a tetrahedron in the Bloch sphere is equivalent to them having linearly independent Stokes vectors. Let's focus on Alice's states, since the states are qubits, they can be expressed in terms of Stokes parameters:
\begin{equation}
    p^{i,x} \rho_A^{i,x} = \frac{1}{2}\sum_{r=0}^{3}P_r^{i,x} \sigma_{r}
\end{equation}
where $\sigma_0$ is the identity and $\sigma_r$, $r=1,2,3$ are the Pauli matrices, while $P_r^{i,x} = p^{i,x} Tr(\sigma_{r} \rho_A^{i,x})$ form the elements of the Stokes vector $\vec{P}^{i,x}$ for that state. Thus:
\begin{equation}
    \text{vec}(p^{i,x} \rho_A^{i,x}) = \frac{1}{2}\sum_{r=0}^{3} P_r^{i,x} \text{vec}(\sigma_{r})
\end{equation}

It is easy to show that $\text{vec}(\sigma_r)^T\text{vec}(\sigma_{r'})=\delta_{rr'}$, which means the inner product of any two $\text{vec}(p^{i,x} \rho_A^{i,x})$ is related to the inner product of the Stokes vectors by a constant factor:
\begin{equation}
\begin{split}
    \text{vec}^T(p^{i,x} \rho_A^{i,x})\text{vec}(p^{i',x'} \rho_A^{x'i'}) =& \frac{1}{2}\sum_{r=0}^{3} P_r^{i,x}P_r^{x',i'}\\
    =& \frac{1}{2} \vec{P}^{i,x}\cdot\vec{P}^{x',i'}
\end{split}
\end{equation}
Thus, since the inner product structure of the rows of $\hat{\gamma}$ is identical to that of the Stokes vectors up to a factor, the linear independence of the Stokes vectors is equivalent to the invertibility of $\hat{\gamma}$.

\section{The virtual picture and optimization of the key rate with semidefinite programming}

In the main text, we provided an overview of how to determine the optimal virtual picture for our protocol using a virtual twisting operation \cite{twisted_eph} and semidefinite programming. Here we provide the full mathematical details of our analytical and numerical techniques.

\subsection{Moving to a virtual picture}
Given states of the form in Eq. \ref{eq:noisy}, we can define a virtual purified picture for the key generation states:
\begin{equation}
       \{p^{0,x}\rho_{A}^{0,x} q^{0,y}\sigma_{B}^{0,y}\} \rightarrow |\zeta\rangle_{\bar{A}\bar{B}A'B'AB}= \sum_{x,y}|x,y\rangle_{\bar{A}\bar{B}}\sum_{m,n = H}^{V} |\gamma^{x,y}_{m,n}\rangle_{A'B'}|m,n\rangle_{AB}
\end{equation}
where $Tr_{\bar{A}\bar{B}A'B'}(\ket{x,y}\bra{x,y}_{\bar{A}\bar{B}}\ket{\zeta}\bra{\zeta}_{\bar{A}\bar{B}A'B'AB})=p^{0,x}\rho_{A}^{0,x} q^{0,y}\sigma_{B}^{0,y}$ fixes the constraint:
\begin{equation}\label{eq:anc_state}
    \braket{\gamma^{x,y}_{m',n'}|\gamma^{x,y}_{m,n}}_{A'B'} \equiv  p^{0,x} q^{0,y} c_{m,m'}^{0,x} d_{n,n'}^{0,y} =\hat{\gamma}_{ts}
\end{equation}
where $\hat{\gamma}_{ts}$ is the same matrix element as Eq. \ref{eq:vec_AB}.

The constraint in Eq. \ref{eq:anc_state} on the ancillary systems is not sufficient to identically fix the purification. Rather, any unitary twisting operation of the form:
\begin{equation}\label{eq:twist_op2}
    U_{\bar{A}\bar{B}A'B'} = \sum_{x,y=0}^{1}|x,y\rangle\langle x,y|_{\bar{A}\bar{B}}\otimes U_{A'B'}^{x,y}
\end{equation}
preserves the real signal states, since:
\begin{equation}
\begin{split}
    &Tr_{\bar{A}\bar{B}A'B'}(\ket{x,y}\bra{x,y}_{\bar{A}\bar{B}}U_{\bar{A}\bar{B}A'B'}\ket{\zeta}\bra{\zeta }_{\bar{A}\bar{B}A'B'AB}U^\dagger_{\bar{A}\bar{B}A'B'})\\
    =& Tr_{\bar{A}\bar{B}A'B'}(\ket{x,y}\bra{x,y}_{\bar{A}\bar{B}}U_{A'B'}^{x,y}\ket{\zeta}\bra{\zeta }_{\bar{A}\bar{B}A'B'AB}U_{A'B'}^{x,y\,\dagger})\\
    =& p^{0,x}\rho_{A}^{0,x} q^{0,y}\sigma_{B}^{0,y}
\end{split}
\end{equation}
Since the produced signal states are independent of this twisting operation, it cannot affect any of the real detection probabilities observed in the execution of the protocol which depend only on the $A,B$ systems. Thus, the characterization of Eve's Gramian matrix elements $\braket{e_{m',n'}^P|e_{m,n}^P}_E$ is independent of the twisting operation. Moreover, since the produced signal states are independent of the twisting operation it never needs to actually be implemented in the real protocol, and can remain a useful virtual analytical tool in the characterization of security after the real signal exchange. Finally, the twisting operation can be chosen after the detection statistics are produced, which gives Alice and Bob the power to adjust their virtual strategy based on what they observe.

The major change when adding the twisting operation is the \textit{definition} of the phase error rates, i.e. how we use the information about Eve's state from $\vec{e}_s$. Consider the joint state between Alice, Bob, their ancillae, and Eve in the purified picture after they apply a twisting operation and send the systems $A,B$ to Eve:
\begin{equation}\label{eq:joint_state_eve}
        U_{\bar{A}\bar{B}A'B'}\ket{\zeta}_{\bar{A}\bar{B}A'B'AB}\rightarrow |\Gamma (U)\rangle_{\bar{A}\bar{B}A'B'EZ}= \sum_{x,y}|x,y\rangle_{\bar{A}\bar{B}}\sum_{m,n = H}^{V} U_{A'B'}^{x,y}|\gamma^{x,y}_{m,n}\rangle_{A'B'}\sum_{z = P}^{F}|e_{m,n}^{z}\rangle_{E}|z\rangle_Z
\end{equation}
Thus, taking the target virtual Bell state to be $\ket{\Phi^+}\bra{\Phi^+}_{\bar{A}\bar{B}}$, the phase error rates now become the \emph{twisted phase error rates}:
\begin{equation}\label{eq:eph_twist}
\begin{split}
    e_X(U)  &= \frac{1}{p_{det}^{0,0}}\langle\Gamma(U)|\left[\right.(|\Psi^-\rangle\langle \Psi^-|
    +|\Phi^-\rangle\langle \Phi^-|)_{\bar{A}\bar{B}}\otimes|P\rangle\langle P|_Z \left.\right] |\Gamma(U)\rangle_{\bar{A}\bar{B}A'B'EZ}\\
    &= \frac{1}{2}-\frac{1}{p_{det}^{0,0}}\sum_{m,n,m',n'} Re\Big[\big(\braket{\gamma_{m',n'}^{0,0}|{U^{0,0}_{A'B'}}^\dagger U^{1,1}_{A'B'}|\gamma_{m,n}^{1,1}}_{A'B'}+\braket{\gamma_{m',n'}^{0,1}|{U^{0,1}_{A'B'}}^\dagger U^{1,0}_{A'B'}|\gamma_{m,n}^{1,0}}_{A'B'}\big)\braket{e_{m',n'}^P|e_{m,n}^P}_E\Big]\\
    \\
    e_Y(U)  &= \frac{1}{p_{det}^{0,0}}\langle\Gamma(U)|\left[\right.(|\Phi^-\rangle\langle \Phi^-|
    +|\Psi^+\rangle\langle \Psi^+|)_{\bar{A}\bar{B}}\otimes|P\rangle\langle P|_Z \left.\right] |\Gamma(U)\rangle_{\bar{A}\bar{B}A'B'EZ}\\
    &= \frac{1}{2}-\frac{1}{p_{det}^{0,0}}\sum_{m,n,m',n'} Re\Big[\big(\braket{\gamma_{m',n'}^{0,0}|{U^{0,0}_{A'B'}}^\dagger U^{1,1}_{A'B'}|\gamma_{m,n}^{1,1}}_{A'B'}-\braket{\gamma_{m',n'}^{0,1}|{U^{0,1}_{A'B'}}^\dagger U^{1,0}_{A'B'}|\gamma_{m,n}^{1,0}}_{A'B'}\big)\braket{e_{m',n'}^P|e_{m,n}^P}_E\Big]
\end{split}
\end{equation}
In general, there exists an optimal purification provided by some $U_{\bar{A}\bar{B}A'B'}$ such that the key rate is maximized. As it is unlikely to choose this optimal purification at random when constructing the problem, we expect the calculation of $e_X(U)$ and $e_Y(U)$ to benefit from an optimization over $U_{\bar{A}\bar{B}A'B'}$, and thus, in general, for the key rate to increase by employing an optimal twisted phase error rate.

\textbf{Remark:} employing noisy, i.e. mixed, signal states in the protocol begs the question of how best to purify the states in the virtual protocol when defining the phase error rate. This necessarily leads to the definition of a twisted phase error rate and a search for an optimal twisting operation with respect to the key rate. The twisting operation is entirely in the virtual picture, so the optimal $U_{\bar{A}\bar{B}A'B'}$ can and should be computed after the exchange of signals and observation of detection probabilities and bit error rates. Although the form of the optimal twisting operation can even be nonlocal across $\bar{A}\bar{B}$ in practice, it does not need to be implemented in the real protocol, so its locality does not matter.

It is worth emphasizing that employing a twisting operation is an optional step in the security proof. Indeed, any phase error rates of the form Eq. \eqref{eq:eph_twist} will supply a suitable lower bound on the key rate, since Eve does not have control over the $A'B'$ systems. Nonetheless, optimizing over the twisting operation will in general boost the key rate, safeguarding against a poor, overly pessimistic initial choice of purification.

We next demonstrate how optimizing the phase error rates over twisting operations can be framed as two semidefinite programs, closing a gap in the previous literature on twisting operations \cite{private_states2}.

\subsection{Semidefinite programs for evaluating the six state key rate}\label{subsec:SDP}
A general optimization problem is of the form \cite{Boyd}:
\begin{center}
\begin{tabular}{ l l }
\texttt{minimize} & $f_0(\textbf{x})$  \\ 
\texttt{s.t.}  & $f_i(\textbf{x})\geq b_i,\ i=1,\dots,m$\\
\end{tabular}
\end{center}
where $\textbf{x}=(x_1,\dots,x_n)$ are the variables over which we optimize; $f_0:\mathbb{R}^n\rightarrow\mathbb{R}$ is the objective function; $f_i:\mathbb{R}^n\rightarrow\mathbb{R}$ are the constraint functions; and, $b_i$ are the constraint bounds. An optimal solution, $\textbf{x}^\star$ would mean that for all $\textbf{z}$ such that $f_i(\textbf{z})\geq b_i$ then $f_0(\textbf{z})\geq f_0(\textbf{x}^\star)$.

Semidefinite programs (SDPs) are a class of convex optimization problems with linear objective and constraint functions over a cone of positive semidefinite (PSD) matrices \cite{Boyd}. That is, the optimization variables, $\textbf{x}$, form the elements of a matrix with non-negative eigenvalues, and $f_i(x) = \textbf{c}_i\cdot\textbf{x}$. They have become an incredibly versatile tool for QKD security proofs in recent years \cite{Coles2016,Winick2018reliablenumerical,Islam2018,Wang2019,Primaatmaja2019,Islam2019}.

\subsubsection{Objective functions}
At first glance, the optimization problem required for the six-state key rate in Eq. 7 of the main text looks daunting. It appears we have two quantities to optimize with the twisting operation, $e_X(U)$ and $e_Y(U)$, appearing in a nonlinear function due to the binary entropy. However, consider a simple change of variable so that the two unknowns are given by:
\begin{equation}
    e_{-}(U) = (e_X - e_Y)(U),\;e_{+}(U) = (e_X + e_Y)(U)
\end{equation}
These remain linear objective functions of the only free variables in the problem $\big\{\braket{\gamma_{m',n'}^{x',y'}|{U^{x',y'}_{A'B'}}^\dagger U^{x,y}_{A'B'}|\gamma_{m,n}^{x,y}}_{A'B'}\big\}$:
\begin{equation}\label{eq:eplus}
    \begin{split}
        e_+(U) &= 1 - \frac{2}{p_{det}^{0,0}}
        \sum_{m,n,m',n'}Re\left(\braket{\gamma_{m',n'}^{0,0}|{U^{0,0}_{A'B'}}^\dagger U^{1,1}_{A'B'}|\gamma_{m,n}^{1,1}}_{A'B'}\braket{e_{m',n'}^P|e_{m,n}^P}_E\right)\\
        &= e_+(U^{0,0\,\dagger}_{A'B'}U^{1,1}_{A'B'})
    \end{split}
\end{equation}
and:
\begin{equation}\label{eq:eminus}
    \begin{split}
        e_-(U) &= - \frac{2}{p_{det}^{0,0}}
        \sum_{m,n,m',n'}Re\left(\braket{\gamma_{m',n'}^{0,1}|{U^{0,1}_{A'B'}}^\dagger U^{1,0}_{A'B'}|\gamma_{m,n}^{1,0}}_{A'B'}\braket{e_{m',n'}^P|e_{m,n}^P}_E\right)\\
        &= e_-(U^{0,1\,\dagger}_{A'B'}U^{1,0}_{A'B'})
    \end{split}
\end{equation}

The only free parameters over which we can optimize are the twisting unitaries, $\{U^{0,1}_{A'B'},U^{1,0}_{A'B'},U^{0,0}_{A'B'},U^{1,1}_{A'B'}\}$, where each of the four unitaries can be defined independently of the others. Here, we have found that the two objective functions in the key rate $e_\pm(U)$, are functions of independent variables: $e_+(U)$ only depends on $U_+ = U^{0,0\,\dagger}_{A'B'}U^{1,1}_{A'B'}$ and $e_-(U)$ only depends on $U_- = U^{0,1\,\dagger}_{A'B'}U^{1,0}_{A'B'}$. This is very good, since it means the difficult task of nonlinear optimization of the six state key rate formula can be avoided. Using the monotonicity of the binary entropy, we can directly optimize $e_\pm(U_\pm)$ within the binary entropy functions:
\begin{equation}
\begin{split}
        R &= \max_{U}p_{det}^{0,0}\left(1-h_2(e_Z)- e_Z h_2\left[\frac{1+e_-(U)/e_Z}{2}\right] - (1-e_Z)h_2\left[\frac{1-[e_+(U)+e_Z]/2}{1-e_Z}\right]\right)\\
        &= p_{det}^{0,0}\left(1-h_2(e_Z)- e_Z h_2\left[\frac{1+\max_{U_-}e_-(U_-)/e_Z}{2}\right] - (1-e_Z)h_2\left[\frac{1-[\min_{U_+}e_+(U_+)+e_Z]/2}{1-e_Z}\right]\right)
\end{split}
\end{equation}
with the extra conditions $0\leq e_-(U_-)\leq e_Z$ and $e_Z\leq e_+(U_+) \leq 1$ so that the arguments of the binary entropy functions remain between 0 and 1.

\subsubsection{Two independent semidefinite programs}
We recall that $\braket{e_{m',n'}^P|e_{m,n}^P}_E$ are known from Eq. \ref{eq:solving_Eve}, while $\braket{\gamma_{m',n'}^{x',y'}|{U^{x',y'}_{A'B'}}^\dagger U^{x,y}_{A'B'}|\gamma_{m,n}^{x,y}}_{A'B'}$ are the optimization variables. This leads to two independent semidefinite programs.

\begin{itemize}
    \item For the linear objective function $e_-(U_-)$, we note the optimization variables
    \begin{equation}\label{eq:emin_gram}
        \braket{\gamma_{m',n'}^{x,(x+1\bmod2)}|{U^{x,(x+1\bmod2)}_{A'B'}}^\dagger U^{y,(y+1\bmod2)}_{A'B'}|\gamma_{m,n}^{y,(y+1\bmod2)}}_{A'B'}
    \end{equation}
    form the $8\times8$ positive semidefinite Gram matrix for the vectors $\{U^{0,1}_{A'B'}\ket{\gamma_{m,n}^{0,1}}_{A'B'},U^{1,0}_{A'B'}\ket{\gamma_{m,n}^{1,0}}_{A'B'}\}$, subject to the eight linear constraints from Eq. \ref{eq:anc_state}:
    \begin{equation}\label{eq:emin_cons}
        \braket{\gamma_{m',n'}^{x,(x+1\bmod2)}|{U^{x,(x+1\bmod2)}_{A'B'}}^\dagger U^{x,(x+1\bmod2)}_{A'B'}|\gamma_{m,n}^{x(x+1\bmod2)}}_{A'B'}=\braket{\gamma_{m',n'}^{x,(x+1\bmod2)}|\gamma_{m,n}^{x,(x+1\bmod2)}}_{A'B'}
    \end{equation}
    The optimization is additionally constrained by $0\leq e_-(U_-)\leq e_Z$.

    \item For the linear objective function $e_+(U_+)$, we note the optimization variables
    \begin{equation}\label{eq:eplus_gram}
        \braket{\gamma_{m',n'}^{x,x}|{U^{x,x}_{A'B'}}^\dagger U^{y,y}_{A'B'}|\gamma_{m,n}^{y,y}}_{A'B'}
    \end{equation}
    form the $8\times8$ PSD Gram matrix for the vectors $\{U^{0,0}_{A'B'}\ket{\gamma_{m,n}^{0,0}}_{A'B'},U^{1,1}_{A'B'}\ket{\gamma_{m,n}^{1,1}}_{A'B'}\}$, subject to the eight linear constraints from Eq. \ref{eq:anc_state}:
    \begin{equation}\label{eq:eplus_cons}
        \braket{\gamma_{m',n'}^{x,x}|{U^{x,x}_{A'B'}}^\dagger U^{x,x}_{A'B'}|\gamma_{m,n}^{x,x}}_{A'B'}=\braket{\gamma_{m',n'}^{x,x}|\gamma_{m,n}^{x,x}}_{A'B'}
    \end{equation}
     The optimization is additionally constrained by $e_Z\leq e_+(U_+) \leq 1$.
\end{itemize}

With that, we have two independent semidefinite programs which can be used to optimize the six state key rate formula. In section \ref{sec:pseudocode}, we provide a pseudocode overview of our numerical technique for calculating the key rate.

\section{$(\delta,p)$-model for signal states}
We consider the following two-parameter $(\delta,p)$-model for the initial states which Alice and Bob prepare:
\begin{equation}\label{eq:qubit_eg}
    \begin{split}
        \rho_{A}^{0,0} = \sigma_B^{0,0} &= (1-p)\ket{\xi^\delta_{00}}\bra{\xi^\delta_{00}} + p\id\\
        \rho_{A}^{0,1} = \sigma_B^{1,0} &= (1-p)\ket{\xi^\delta_{01}}\bra{\xi^\delta_{01}} + p\id\\
        \rho_{A}^{1,0} = \sigma_B^{0,1} &= (1-p)\ket{\xi^\delta_{10}}\bra{\xi^\delta_{10}} + p\id\\
        \rho_{A}^{1,1} = \sigma_B^{1,1} &= (1-p)\ket{\xi^\delta_{11}}\bra{\xi^\delta_{11}} + p\id\\
    \end{split}
\end{equation}
where the states $\ket{\xi^\delta}$ are of the form:
\begin{equation}\label{eq:mod_error}
    \begin{split}
        \ket{\xi^\delta_{00}} &= \ket{H}\\
        \ket{\xi^\delta_{01}} &= -\sin\frac{\delta}{2}\ket{H}+\cos\frac{\delta}{2}\ket{V}\\
        \ket{\xi^\delta_{10}} &= \cos\frac{\pi+\delta}{4}\ket{H}+\sin\frac{\pi+\delta}{4}\ket{V}\\
        \ket{\xi^\delta_{11}} &= \cos\frac{-\pi+\delta}{4}\ket{H}+i\sin\frac{-\pi+\delta}{4}\ket{V}\\
    \end{split}
\end{equation}
The states $\ket{\xi^\delta}$ parametrized by $\delta$ are a model for Alice and Bob attempting to prepare $\{\ket{H},\ket{V},\nicefrac{\ket{H}+\ket{V}}{\sqrt{2}},\nicefrac{\ket{H}-i\ket{V}}{\sqrt{2}}\}$, but each state is subject to a different, constant state-dependent modulation error. The pure $\ket{\xi^\delta}$ states and the resulting key rates were considered in the loss-tolerant protocol \cite{loss_tol}. Additionally, were the modulation error a random variable subject to a distribution on the Bloch sphere, we expect the average state to be mixed with a shorter than unit Bloch vector. This effect is accounted for with the depolarizing channel parametrized by $p$, which indicates with some probability the maximally mixed state is sent, shortening the Bloch vector. The depolarizing channel can also be used to model any thermal photons that are accidentally produced during state preparation.

\section{Pseudocode for key rate calculation}\label{sec:pseudocode}
Here we present a sketch of our numerical implementation for calculating key rates. For the semidefinite programs we employed CVXPY \cite{cvxpy,cvxpy_rewriting}, a convex optimization library for Python.  All codes are available upon request.

\begin{algorithm}[H]
\caption{Key rate function}
\begin{algorithmic}
\Function{keyrate}{$\rho_A,\sigma_B,p_A,q_B,p_{dark},\eta,l$} 

    \# $\rho_A$ and $\sigma_B$ are lists of Alice and Bob's four density matrices
    
    \# $p_A$ and $q_B$ are lists of the probabilities for sending their four states
    
    \# $p_{dark}$ is the dark count probability per detector
    
    \# $\eta$ is the overall transmissivity
    
    \# $l$ is the Alice-Charlie distance (same for Bob)
    
    \#
    
    \# Probability of losing a photon
    
    $p_0 = 1-\eta10^{-0.2l/20}$

    \# Extract protocol statistics
    
    $\vec{p}_{det}$,$\hat{\gamma}$= stats($\rho_A$,$\sigma_B$,$p_A$,$q_B$,$p_0$,$p_{dark}$)
    
    \# Key generation detection probability
    
    $p_{det}^{0,0} = \sum_{i=0}^3 \vec{p}_{det}[i]$
    
    \# Bit error rate
    
    $e_Z = \vec{p}_{det}[1]+\vec{p}_{det}[2]$
    
    \# Solving for Eve's Gramian matrix (Eq. 3 of main text)
    
    $\vec{e} = \hat{\gamma}^{-1} \vec{p}_{det}$
    
    \# Phase error rates
    
    $e_{-} = \text{emin}(\rho_A,\sigma_B,\vec{e},e_Z,p_{det}^{0,0})$
    
    $e_{+} = \text{eplus}(\rho_A,\sigma_B,\vec{e},e_Z,p_{det}^{0,0})$
    
    \# Key rate
    
    $R = p_{det}^{0,0}\left[1 - h_2(e_Z) - e_Z h_2\left(\frac{1+e_-/e_Z}{2}\right) - (1-ez)h_2\left(\frac{1-(e_++e_Z)/2}{1-e_Z}\right)\right]$
    
    \textbf{return} R
    
\EndFunction

\end{algorithmic}
\end{algorithm}

\begin{algorithm}[H]
\caption{Protocol statistics function}
\begin{algorithmic}
\Function{stats}{$\rho_A$,$\sigma_B$,$p_A$,$q_B$,$p_0$,$p_{dark}$}

\# Loop over all 16 combinations of states in lists: i,j=1,2,3,4

\# Probability of passing if both photons arrive

$p_{pass}[4i+j] = p_A[i]q_B[j]Tr(\rho_A[i]\sigma_B[j] \ket{\Phi^+}\bra{\Phi^+}_{AB})$

\# Detection probability including dark counts and loss

$p_{det}[4i+j] = (1-p_0)^2 (1-p_{dark})^2 p_{pass}[4i+j] + 2p_A[i]q_B[j][p_0^2p_{dark}^2(1-p_{dark})^2 + p_0(1-p_0)p_{dark}(1-p_{dark})^2]$ 

\# Filling the 16 rows of the $\hat{\gamma}$ matrix

$\hat{\gamma}[4i+j,:] = p_A[i]q_B[j]\text{vec}(\rho_A[i]\sigma_B[j])$ 

\textbf{return} $\vec{p}_{det}, \hat{\gamma}$
\EndFunction
\end{algorithmic}
\end{algorithm}

\begin{algorithm}[H]
\caption{Phase errors}
\begin{algorithmic}

\Function{emin}{$\rho_A,\sigma_B,\vec{e},e_Z,p_{det}^{0,0}$}

\# Reshape $\vec{e}$ into a matrix

$\hat{e} = \text{reshape}(\vec{e})$

\# We use the CVXPY and Mosek packages for solving semidefinite programs

import cvxpy

\# Define the 8x8 Gramian matrix from Eq. \ref{eq:emin_gram} for the $A'B'$ systems as the optimization variables of the system

$G =$ cvxpy.Variable((8,8))

\# Define a list of constraints on G, such as PSD and constraint from Eq. \ref{eq:emin_cons}

constraints = [$G\succeq 0$]

constraints += $\left[G[4i+2m+n,4i+2m'+n']=\rho_A[i][m,m']\sigma_B[(i+1)\bmod2][n,n']\right]$ \#$i=0,1;m,m',n,n'=0,1,2,3$

\# Define the objective function

$e_{-} = -\frac{2}{p_{det}^{0,0}}\sum_{m,m',n,n'} Re(\hat{e}[2m+n,2m'+n']G[2m+n,4+2m'+n'])$

constraints += $[e_{-}\geq0,e_{-}\leq e_Z]$

\# Use cvxpy to solve problem

prob = cvxpy.Problem(cvxpy.Maximize($e_{-}$),constraints)

prob.solve(solver = cvxpy.MOSEK)

$e_{-}$ = prob.value

\textbf{return}  $e_{-}$
\EndFunction
\\
\Function{eplus}{$\rho_A,\sigma_B,\vec{e},e_Z,p_{det}^{0,0}$}

\# Reshape $\vec{e}$ into a matrix

$\hat{e} = \text{reshape}(\vec{e})$

\# We use the CVXPY and Mosek packages for solving semidefinite programs

import cvxpy

\# Define the 8x8 Gramian matrix from Eq. \ref{eq:eplus_gram} for the $A'B'$ systems as the Variable of the system

$G =$ cvxpy.Variable((8,8))

\# Define a list of constraints on G, such as PSD and constraint from Eq. \ref{eq:eplus_cons}

constraints = [$G\succeq 0$]

constraints += $\left[G[4i+2m+n,4i+2m'+n']=\rho_A[i][m,m']\sigma_B[i][n,n']\right]$ \#$i=0,1;m,m',n,n'=0,1,2,3$

\# Define the objective function

$e_{+} = 1-\frac{2}{p_{det}^{0,0}}\sum_{m,m',n,n'} Re(\hat{e}[2m+n,2m'+n']G[2m+n,4+2m'+n'])$

constraints += $[e_{+}\geq e_Z,e_{+}\leq 1]$

\# Use cvxpy to solve problem

prob = cvxpy.Problem(cvxpy.Minimize($e_{+}$),constraints)

prob.solve(solver = cvxpy.MOSEK)

$e_{+}$ = prob.value

\textbf{return}  $e_{+}$
\EndFunction
\end{algorithmic}
\end{algorithm}

\bibliography{biblio_supp}